\documentclass[preprint]{aastex}

\shorttitle{A Machine Learning Classifier for Fast Radio Bursts}
\shortauthors{K.~L. Wagstaff et al.}

\usepackage{natbib}
\usepackage{color}
\usepackage{subfigure}
\usepackage{todonotes}
\usepackage{booktabs}
\usepackage{graphicx}
\usepackage[normalem]{ulem}

\long\def\comment#1{}

\newcommand{\edit}[2]{#2}

\begin{document}

\title{A Machine Learning Classifier for \edit{}{Fast} Radio
  \edit{Transient}{Burst} Detection at the VLBA} 

\author{Kiri L. Wagstaff\altaffilmark{1}, 
  Benyang Tang\altaffilmark{1}, 
  David R. Thompson\altaffilmark{1},  
  Shakeh Khudikyan\altaffilmark{1},   
  Jane Wyngaard\altaffilmark{1},      
  Adam T. Deller\altaffilmark{2}, 
  Divya Palaniswamy\altaffilmark{3,4}, 
  Steven J. Tingay\altaffilmark{3,5}, 
  and Randall B. Wayth\altaffilmark{3}} 

\email{kiri.l.wagstaff@jpl.nasa.gov}

\altaffiltext{1}{Jet Propulsion Laboratory, California Institute of Technology, 4800 Oak Grove Drive, Pasadena, CA 91109, USA }
\altaffiltext{2}{ASTRON, the Netherlands Institute for Radio Astronomy, Oude
Hoogeveensedijk 4, 7991 PD Dwingeloo, The Netherlands}
\altaffiltext{3}{ICRAR, Curtin University, Bentley, WA 6845, Australia}
\altaffiltext{4}{University of Nevada, 4505 S. Maryland Parkway, Las
  Vegas, NV 89154, USA }
\altaffiltext{5}{ARC Centre of Excellence for All-Sky Astrophysics (CAASTRO)}

\begin{abstract}

Time domain radio astronomy observing campaigns frequently generate
large volumes 
of data.  Our goal is to develop automated methods that can identify
events of interest buried within the larger data stream.  The V-FASTR
fast transient system was designed to detect rare fast radio bursts
(FRBs) within data collected by the Very Long Baseline Array.  The
resulting event candidates constitute a significant burden in
terms of subsequent human reviewing time.  We have trained and
deployed a machine learning classifier that marks each candidate
detection as a pulse from a known pulsar, an artifact due to radio
frequency interference, or a potential new discovery.  The classifier
maintains high reliability by restricting its predictions to those
with at least 90\% confidence.  We have also implemented several
efficiency and usability improvements to the V-FASTR web-based
candidate review system.  Overall, we found that time spent reviewing
decreased and the fraction of interesting candidates increased.  The
classifier now classifies (and therefore filters) 80-90\% of
the candidates, with an accuracy greater than 98\%, leaving only the
10-20\% most promising candidates to be reviewed by humans.

\end{abstract}



\section{Introduction}

\edit{}{The Very Long Baseline Array (VLBA) consists of 10 widely
separated radio antennas.  Their locations range from Mauna Kea,
Hawaii, to St. Croix in the Virgin Islands, yielding baselines of up
to 8000 km.  Data is collected individually by each 25\,m antenna
and then sent to the correlator in Socorro, NM.  Excellent time and
angular resolution enable precise detection and localization of
coincident signals observed by multiple antennas.}

Fast radio bursts
(FRBs)~\citep{Lorimer2007,Thornton2013,burke-spolaor:frb14}
\edit{}{are radio phenomena of particular interest}.
These are non-repeating, 
short-duration (millisecond to sub-millisecond), broad-band radio
pulses that are observed on Earth with a frequency-dependent 
time of arrival.  The amount of {\em dispersion}, or the delay between the
arrival of the signal at the highest and lowest frequencies observed,
is dependent on the path length of the signal through ionized plasma (and
its density) between the source and the observer. Highly dispersed
transient events are of particular interest 
since they may have an extragalactic origin.  Potential sources of
these events include gamma-ray bursts (GRBs)
\edit{(Cameron et al. 2005) and rotating radio transients
  (RRATs) (McLaughlin et al. 2006).}{\citep{zhang:frb-grb14},   
supra-massive neutron stars~\citep{falcke:frb-neutronstar14}, 
binary neutron star mergers~\citep{totani:frb-binary13}, 
binary white dwarf mergers~\citep{kashiyama:frb-whitedwarf13}, 
flaring stars~\citep{loeb:frb-flare14}, and many more.} 

In addition to
astronomical signals, local (Earth-based) sources of interference
can generate intrinsically dispersed signals that in some ways
resemble FRBs.  One prominent example was the discovery of the
so-called ``Perytons''~\citep{Burke2011} that ultimately were
determined to have been 
generated by a microwave oven near the Parkes radio
telescope~\citep{petroff:perytons15}.  Even within sources  
categorized as astronomical FRBs, there is considerable variety: the
prototypical Lorimer Burst~\citep{Lorimer2007} is far brighter than
any of the subsequently discovered bursts, for instance.
It would significantly improve our understanding of these events to
localize an incoming pulse through wide baseline interferometry.
However, this would require re-analyzing the data after a pulse is
identified, demanding archival storage of an infeasible quantity of
raw data.

To better understand the variety and incidence of FRBs, we seek to
detect and catalog as many such events as possible.
The ideal transient detection system would be able to detect FRBs
across a wide range of frequencies, produce high time-resolution data,
discriminate against RFI, have good sky coverage, and localize the
burst on the sky with enough angular resolution to identify the origin
of the burst. The VLBA is well suited to such a task, and
\edit{V-FASTR was designed to take advantage of its capabilities.}{}
the V-FASTR (VLBA Fast Transient)
system~\citep{Thompson2011,Wayth2011} was created to search \edit{the  
radio sky}{data collected by the VLBA} for \edit{interesting
  transient events known as}{FRBs}.  

\subsection{The V-FASTR System for FRB Detection}

\edit{Most imaging telescopes are sensitive to timescales (usually seconds)
much longer than the millisecond range of transients.}{Searches for
transients in the image domain are usually confined to longer-duration
events ($\sim$seconds or longer), due to the extreme computational
complexity of searching for short and dispersed signals such as single
pulses from radio pulsars or FRBs.  An overview of current and future
image plane transient searches is given by~\cite{fender:frb15}.}
\edit{Pulsar detection uses time averaging to achieve millisecond-scale
sensitivity, but it is}{Searches for fast radio transients use short
  integration times (millisecond or shorter) to preserve sensitivity,
  but they are} usually carried out on large single dishes with
narrow fields of view (FOV), poor angular resolution, and high
sensitivity to RFI.
\edit{}{However, we note that some pilot studies using imaging to
  search for fast radio transients have been carried
  out~\citep{law:frb15}.} 

The VLBA provides a smaller FOV (0.27 deg\textsuperscript{2})
than other similar experiments~\citep{Wayth2011}
with a sensitivity of 0.3 Jy (1.4 GHz). 
As a set of distributed antennas, however, the VLBA offers a
distinct advantage in RFI rejection because signals that are not
observed by multiple antennas can be automatically filtered out.
The VLBA's long baseline (up to 8000 km) enables extremely good
localization of any detected sources (within a few milliarcseconds)
which is invaluable for interpreting and following up on any detections.

Further, the VLBA's flexible DiFX software
correlator~\citep{Deller2011} enables the generation of
short-integration spectrometer data for each antenna at minimal
additional cost as part of the F stage, and because the time and
frequency resolution properties of the spectrometer data are
configurable, it is possible to customize the processing. A 1-ms time
resolution for observing frequencies around 1.4 GHz was chosen for
V-FASTR as a compromise between signal detectability and data volume or
computational effort.
 
V-FASTR was
designed to operate commensally at the \edit{Very Long Baseline Array
(}{}VLBA\edit{)}{}, meaning that it passively analyzes all data
collected by the 
array in support of a variety of imaging campaigns.  This enables 
the potential for FRB detections even in campaigns with other primary
scientific goals.  In 2014, V-FASTR was also granted 700 hours of its
own observing time to conduct a more systematic scan and reduce the
sampling bias induced by observing only those sky locations selected
by investigators for other purposes.

V-FASTR employs a highly efficient, real-time candidate detection
system that processes \edit{all}{} incoming VLBA data and saves out
information about possible FRBs~\citep{Thompson2011,Wayth2011}.
\edit{}{FRB candidates are defined as strong signals that are
  correlated across multiple antennas.  V-FASTR balances sensitivity
  with robustness by adaptively deciding which antennas to use based
  on the current noise environment.}
VFASTR's real-time \edit{}{``triage''} operation permits analysis of
much larger data volumes than would be possible for
\edit{asynchronous}{offline} processing.  While it analyzes a 
majority of VLBA data, it only preserves the tiny fraction of voltage
data associated with candidate events.  \edit{This preserves the VLBA's
critical interferometry advantage with modest storage
requirements.}{} Later, those 
candidates are reviewed by experts to determine whether they originate
from a known source (e.g.,~pulsar), an artifact or radio frequency
interference (RFI), or a truly novel source.
\edit{We developed}{The review process is enabled by} a web-based
classification and review system that 
provides a human-friendly interface for \edit{the process of}{} reviewing
the tens to hundreds of candidates detected per day~\citep{Hart2014}.

To date, no new FRBs have been detected, but the time spent observing
at a range of frequencies and pointing locations has informed the
determination of upper limits on the expected rate of such
events~\citep{Wayth2012,trott:limits13}.  Blind detections of many
known pulsars~\citep{Thompson2013} have validated the system's
sensitivity. V-FASTR serves as a
pathfinder experiment to illustrate how commensal science can be done,
both for added value in current science campaigns and as a way to
scale up to the unprecedented data volumes anticipated for the Square
Kilometre Array~\citep{macquart2014} and other future instruments.

\subsection{Contributions}

In this paper, we describe two new advances to the V-FASTR review
system that were designed to reduce the human effort required to
evaluate the V-FASTR candidates.
%
First, we developed a machine learning classifier that automatically
predicts the correct classification of 80-90\% of the V-FASTR 
candidates, tagging them as created by known pulsars or \edit{radio
frequency interference (}{}RFI\edit{)}{}. By \edit{}{pre-}classifying
a large proportion of 
uninteresting (to this experiment) candidates, the classifier enables
human reviewers to focus their time on the remaining potentially
interesting candidates. \edit{These remaining candidates represent new
phenomena that the classifier cannot confidently classify as pulsars
or RFI.}{}  As each such candidate is considered and tagged with its
appropriate class by human reviewers, the classifier re-trains with
this new information and continually improves its knowledge of radio
transient characteristics.

Second, we implemented several improvements to the V-FASTR web-based
candidate review portal.  This system is vital to support the
geographically distributed team of reviewers.  The interface enables
the fast compilation of summary statistics about the candidates that
have been detected and, as noted above, it also enables continual
improvement for the machine classifier in the form of new feedback
from the reviewers.
We have implemented (1) efficiency improvements that greatly increased
the responsiveness of the system, (2) user interface improvements
identified by a user study, (3) and user authentication to improve
usability (by providing a customized review list for each user) and
enable per-user activity tracking.

Taken together, these advances contribute to a solution to the
data volume challenge involved in VLBA data processing and radio
transient detection.  In addition to rapid data triage by the
real-time detection system, it is vital to minimize the human effort
required to review the candidate events.
The V-FASTR classifier provides a necessary
\edit{first-order filter}{initial step} that 
sets aside candidates that are known to be uninteresting and enables
reviewers to devote their time to the review of the remaining
candidates.  The improved interactive web interface serves the dual
purposes of visualizing the candidates and collecting human
evaluations of each one.   
This solution is significant given the anticipated arrival of
instruments such as the Square Kilometre Array and its predecessors,
which promise to increase the
potential observational parameter space for future radio astronomy
surveys by orders of magnitude~\citep{Lazio2009b}.
%
Going forward, V-FASTR will continue to search for, classify, and send
alerts for promising candidates, potentially leading to new discoveries.

\section{Background and Related Work}

\subsection{Radio Transient Detection Studies}

A number of other studies have detected radio transients using
instruments other than the VLBA.
Data collected by the Parkes radio telescope has been studied
extensively following the detection of the Lorimer
burst~\citep{Lorimer2007}. 
\edit{Keith et al.~(2010) found 5 new millisecond pulsars with
large dispersion measures.}{} 
In an archival survey of Parkes data, \cite{Keane2012} found an FRB
that was observed (but unnoticed) in 2001 and proposed that its source
could be a radio-emitting magnetar. 
\cite{Thornton2013} detected four FRBs in newly collected Parkes data
since 2010,
\cite{burke-spolaor:frb14} reported another FRB in data from 2001,
and \cite{Ravi2015} and~\cite{Petrof2015} each detected one new FRB
during real-time observing in 2013 and 2014, respectively.

\cite{Arecibo2014} reported the first FRB detection on an
instrument other than the Parkes radio telescope (the Arecibo L-Band
Feed Array or ALFA).
\cite{Siemion2011b} searched 450 hours of data from the Allen
Telescope Array (ATA) but did not find any new transients.
\cite{masui2015} found an FRB in archived data from the Green Bank
Hydrogen Intensity Mapping survey.

\begin{figure}[t]
\centerline{\includegraphics[width=0.9\textwidth]{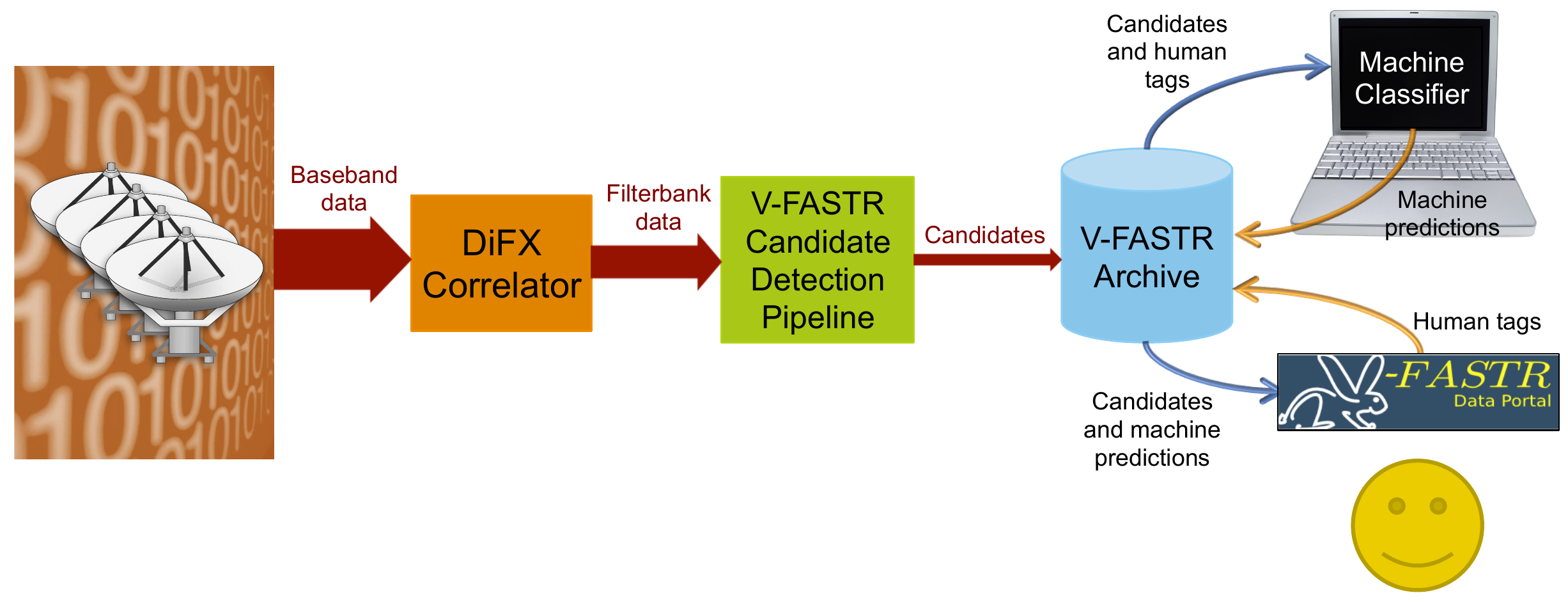}}
\caption{ The V-FASTR system operates commensally with regular VLBA
  data collection and correlation.  Data is transferred from the
  antennas to the DiFX software correlator~\citep{Deller2011}.
  Filterbank data output by the spectrometers is analyzed by the
  real-time candidate detection system~\citep{Thompson2011}, and those
  candidates are stored in the V-FASTR archive.  Data volume is
  reduced at each step.  The candidates are shown to human reviewers
  via the web portal.  Candidates tagged by reviewers are used to
  re-train the classifier, which in turn generates predictions that
  are displayed to reviewers to aid in their review process.}
\label{fig:pipeline}
\end{figure}

\subsection{The V-FASTR Radio Transient Detection and Review System}

Figure~\ref{fig:pipeline} outlines the complete V-FASTR system. Each
VLBA observation can employ up to ten 25-m antennas for simultaneous
observations from different geographic locations.  Once the
observation is complete, hard drives containing the raw (baseband)
data are shipped from each antenna to the Array Operations Center of
the National Radio Astronomy Observatory in Socorro, NM, for
correlation and analysis. The V-FASTR event detection pipeline runs in
real time in parallel with the DiFX correlator.  The correlator
generates filterbank data that captures the observed signal intensity
at each antenna for a range of frequency bins as a function of time.
V-FASTR employs a robust, adaptive summing technique to detect
candidate events in the filterbank data~\citep{Thompson2011}.  The
data associated with each candidate (including, for the top few 
candidates per hour, raw voltage data that would enable localization
to the milliarcsecond level) is saved to disk, and figures are
generated to display the data for human review.  The rest of the data
is discarded, and the hard drives are erased once correlation
completes to enable their re-use for future observations.  Meta-data
associated with each detection, including its timestamp, the array
pointing sky coordinates (right ascension and declination), the signal
strength, and the dispersion measure, are also saved to aid in the
candidate review process.

In order to ensure that no interesting event is missed, the real-time
system uses a lenient threshold that also admits some non-FRB events.
Candidates consist of pulsar pulses, spurious correlated radio
frequency interference (RFI), and other potentially unknown phenomena.
The number of candidates generated by V-FASTR each day ranges from
zero to tens to hundreds to thousands, depending on the observational
target and environmental \edit{}{(temperature and RFI)}
conditions. These candidates are highly 
diverse, and no simple rule can easily recognize or anticipate  all
cases. Consequently, human review is the safest way to filter them
further without missing an interesting event.

We developed a candidate event classifier to reduce the reviewing
burden by automatically tagging events that can be confidently
classified as pulses from a known pulsar or as RFI (artifacts).  The
remaining candidates consist of pulses without a known origin or
explanation; these are the candidates that most require human
assessment and that have the highest chance of containing a new
discovery.   

The V-FASTR review process relies on a web-based review portal that
offers reviewers worldwide access to the events for review and
evaluation~\citep{Hart2014}.  Reviewers, interested PIs, and the
public are able to search and browse the resulting classified
candidates\footnote{\url{http://curta.pawsey.org.au/}}.  Human review
decisions are also used to update and re-train the machine classifier,
so that the system automatically and transparently improves over time.


	



In this paper, we first present the V-FASTR candidate classifier and
describe how it is trained and evaluated
(Section~\ref{sect:classifier}).  Next, we describe the web-based
review portal and our recent improvements and advances
(Section~\ref{sect:portal}).  
Section~\ref{sect:results} shares the results of the classifier's use
in the real-time system and web portal. Finally, 
Section~\ref{sect:conc} describes the system's current use and discusses
how the V-FASTR approach might be used advantageously elsewhere
in the future.


\section{V-FASTR Candidate Classifier for Radio Transients}
\label{sect:classifier}

The V-FASTR candidate classifier analyzes each candidate that is
detected by the real-time system.  It employs a trained random forest
classifier~\citep{breiman:rf01} to predict the class for each new
candidate, then consults a database of known pulsars to further refine
its predictions.  Predictions that are sufficiently confident are
added to the meta-data associated with the candidate and used to
reduce the number of candidates that require human review.

\subsection{V-FASTR Classifier Features}
\begin{figure}
\centerline{\includegraphics[width=4.5in]{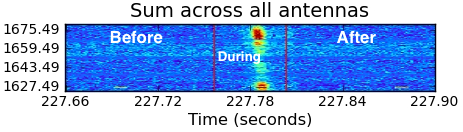}}
\caption{Example V-FASTR candidate with signal strength as a function
  of frequency (in MHz) and time.  Vertical red lines indicate the start and
  end of the event.  ``Before'', ``during'', and ``after''
  regions are used to calculate descriptive features.}
\label{fig:ex}
\end{figure}

The V-FASTR detection system identifies candidates by
de-dispersing the radio frequency data from each observing antenna and
computing a robust sum across antennas to identify correlated
events~\citep{Thompson2011}.  
Filterbank data for each candidate is saved out to disk for further
analysis.  Figure~\ref{fig:ex} shows an example of a candidate event
detected by V-FASTR, in which signal strength is shown as a function
of frequency (y axis) and time (x axis).  Vertical red lines indicate
the start and end of the detected event.  The saved data includes
observations for a buffer period before and after the event to provide
context for the observing conditions at the time that the event
occurred. 

For each candidate, the system constructs a feature vector that
captures key information needed to classify it.  
\edit{}{The features were chosen based on years of manual review
  experience with the types of artifacts and effects the data
  exhibited during that period.  They leverage general image
  statistics of relevant regions in the data stream before, during,
  and after a candidate event.}
The ten features are as follows:
\begin{enumerate}
\item The minimum observing frequency (in Hz).  This is important
  because some bands are more prone to RFI than others.
\item The estimated dispersion measure (DM) (in pc/cc). This feature
  helps exclude non-astrophysical events.
\item The signal to noise ratio (SNR) of the detection.  This
  feature can influence the statistics of many other
  characteristics, so it is important to consider in the aggregate
  decision. 
\item Max asymmetry: The maximum \edit{}{difference} (across frequency
  channels) \edit{difference}{} in observed intensity {\em before} the event versus {\em
    after} the event (see Figure~\ref{fig:ex}).  This attribute is
  often extreme for narrow-band transient RFI.
\item Mean asymmetry: The mean \edit{}{difference} (across frequency channels) 
  \edit{difference}{} in observed intensity {\em before} the event versus {\em
    after} the event.  This feature helps to recognize system state
    changes which can cause step-function changes in the signal level.
\item Max outlierness: The maximum \edit{}{difference} (across
  frequency channels) 
  \edit{difference}{} in observed intensity {\em during} the event versus the
  concatenation of observations {\em before and after} the event.
  This helps to recognize unstable interference conditions over large
  numbers of antennas.
\item Mean outlierness: The mean \edit{}{difference} (across frequency channels)
  \edit{difference}{} in observed intensity {\em during} the event versus {\em
    before and after} the event.
\item Max zeros: The maximum \edit{}{ratio} (across frequency channels) \edit{ratio}{} of
  signal dropouts {\em during} the event versus {\em before} the
  event.  Network dropouts occasionally occur during correlation, and
  they can create confusing null data segments in the time series.  These
  segments appear as simultaneous signals across all antennas, so it
  is important to make an explicit provision to handle them.
\item Mean zeros: The mean \edit{}{ratio} (across frequency channels) \edit{ratio}{} of signal
  dropouts {\em during} the event versus {\em before} the event.
\item The log ratio of covariance across antennas {\em during} the event
  versus {\em before and after} the event.
\end{enumerate}

\subsection{V-FASTR Classification Hierarchy: Pulsars, Artifacts, and
  Good Candidates} 

\begin{figure}[t]
\centerline{\includegraphics[width=4in]{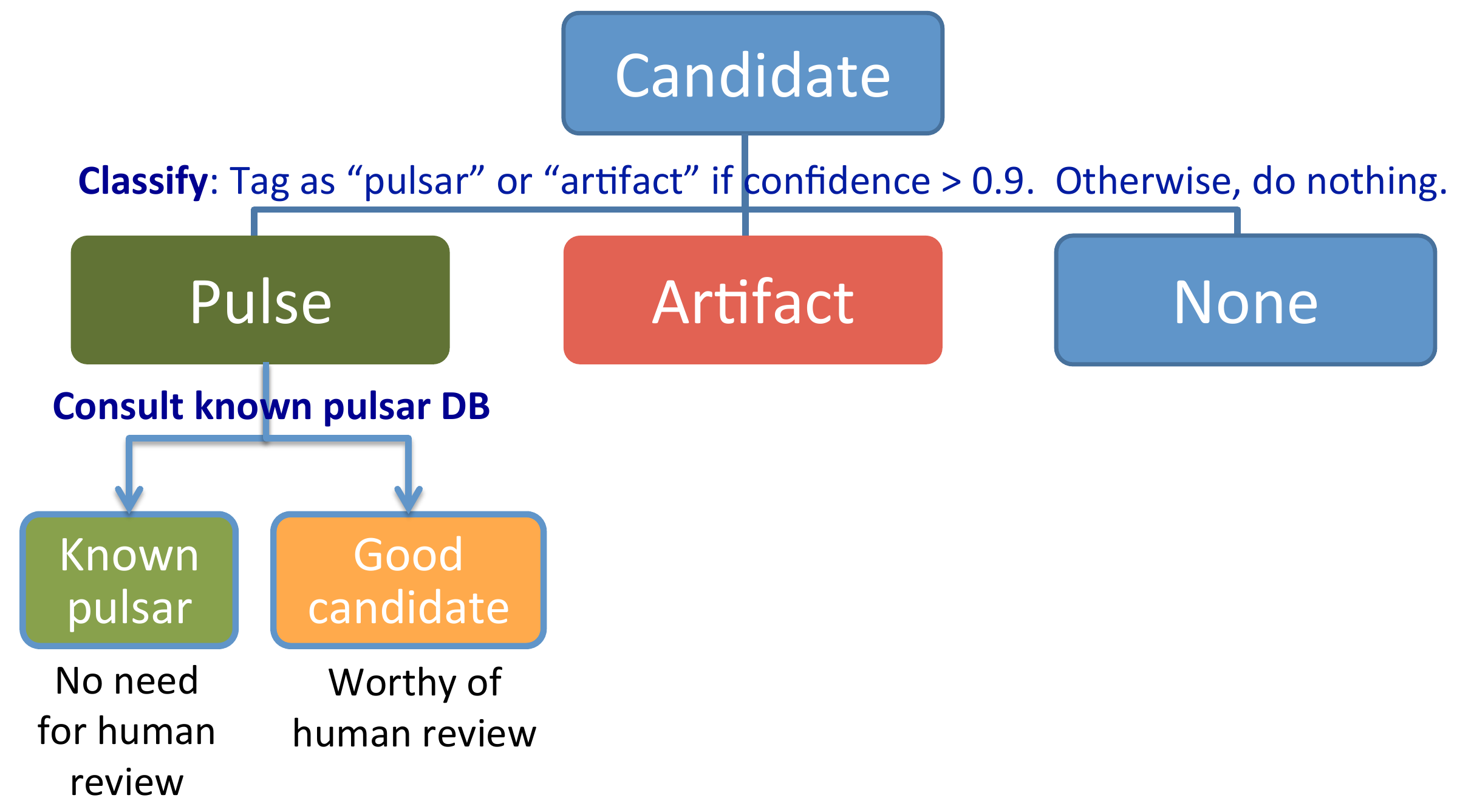}}
\caption{Hierarchical V-FASTR candidate classifier structure.}
\label{fig:cl}
\end{figure}

The V-FASTR classifier adopts a two-stage hierarchical approach to
classifying candidates.  First, \edit{the}{a} random forest classifier is
trained to predict \edit{}{the probability} that a candidate is a
``pulse''\edit{,}{ or} an ``artifact.''  
\edit{or ``none.''  Effectively, it is an abstaining classifier that 
predicts ``none'' (see top level of Figure~\ref{fig:cl}) if it cannot
classify an event as a pulse or artifact with a confidence greater
than $90\%$.}{If the most probable
  class does not have a posterior probability of at least $0.9$, the
  classifier abstains (predicts ``none''; see the top level of
  Figure~\ref{fig:cl}).} 

Candidates classified as ``pulse'' are further refined by consulting
the ATNF Pulsar Catalogue~\citep{manchester:pulsars}, as shown in the
bottom level of Figure~\ref{fig:cl}.  If the array pointing center is
sufficiently close to a known pulsar, and the estimated DM is within
50 pc/cc of the known DM for that pulsar, then the system changes
``pulse'' to ``pulsar.''  If not, the system changes ``pulse'' to
``good candidate''.  These are the candidates that most merit human
review, since they have the characteristics of a pulsar pulse (or FRB)
but do not correspond to a known source.  We define a ``sufficiently
close'' pointing center as one that is within 2 times the full-width
half-maximum (FWHM) beamwidth of the VLBA configuration when the
candidate was detected.  FWHM is $1.22 * \lambda / D$, where $\lambda
= 3.0\times 10^8 / f$, $f$ is the lowest observing frequency, and $D$
is the antenna diameter (25 m).
\edit{}{This two-stage approach to classification reduces the number
  of pulsar database lookups greatly without increasing the complexity
  of the classifier. }

The benefit of the classifier is in reducing the number of candidates
that require human review.  Given sufficiently reliable classifier
predictions, reviewers can prioritize candidates in this order: ``good
candidate'', unclassified candidates (which could represent a new 
phenomenon), ``pulsar'', and ``artifact.''  Therefore, the most
promising candidates will be examined first.

\subsection{V-FASTR Classifier Evaluation}
\label{sect:perf}

Relying on the classifier to tag, sort, and filter candidates requires
that it first demonstrate sufficiently reliable performance.  
We evaluated performance by collecting 7,649 candidates that were
tagged by reviewers as 
``Pulsar'' or ``Artifact'' and conducting 10-fold cross-validation to
assess the classifier's ability to generalize from the training data.
We divided the data set into 10 equally sized ``folds'' and then
repeatedly trained a classifier on nine folds and evaluated its
performance on the held-out tenth.  After doing this 10 times, we had
obtained held-out predictions for all of the labeled data and could
calculate performance in this simulation of prediction on new data.
The classifier had an overall accuracy (agreement with reviewer tags)
of 95.8\%.  

\begin{table}[t]
\caption{Confusion matrix showing the number of candidates classified
  as ``Artifact'' or ``Pulsar'' by the classifier (rows) compared to their
true class labels (columns).  Overall accuracy was 95.8\%.}
\label{tab:cm}
\begin{center}
\begin{tabular}{|l|r|r|r|} \hline
         & \multicolumn{2}{|c|}{True class} & \\ 
         & Artifact & Pulsar & Total \\ \hline
Artifact & 2893     &   199  & 3092 \\
Pulsar   &  124     &  4433  & 4557 \\ \hline
Total    & 3017     &  4632  & 7649 \\ \hline
\end{tabular}
\end{center}
\end{table}

A breakdown of artifact and pulsar classifications on the labeled data
set is shown in the confusion matrix in Table~\ref{tab:cm}.  For
example, the classifier correctly classified 4,433 pulsar candidates.
There were 124 false detections and 199 missed detections, yielding a
false positive rate of 0.03 and a false negative rate of 0.06. 

\begin{figure}[t]
\subfigure[Cross-validated accuracy]{
\includegraphics[width=3.2in]{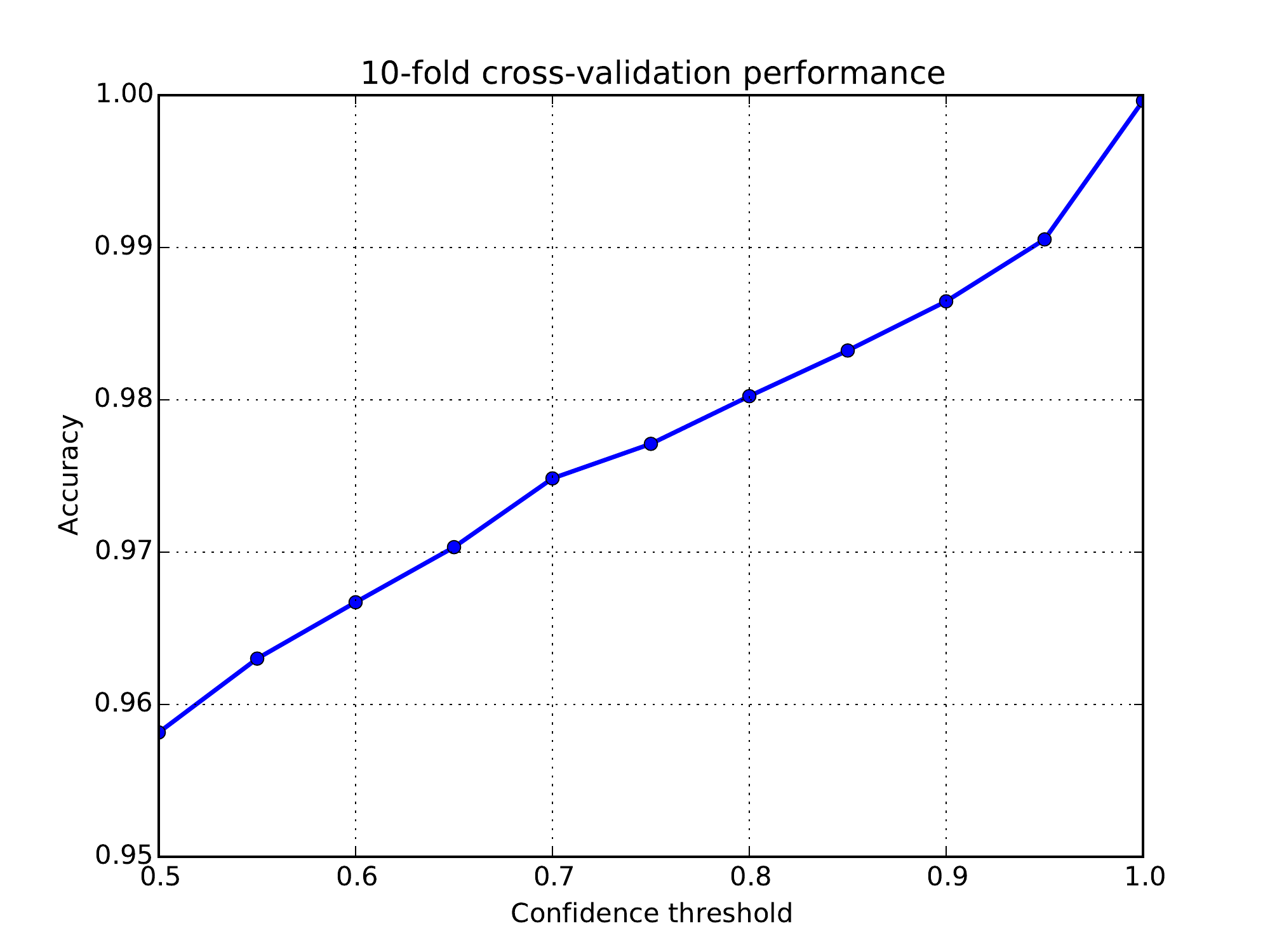}
}
\subfigure[Percent of candidates classified]{
\includegraphics[width=3.2in]{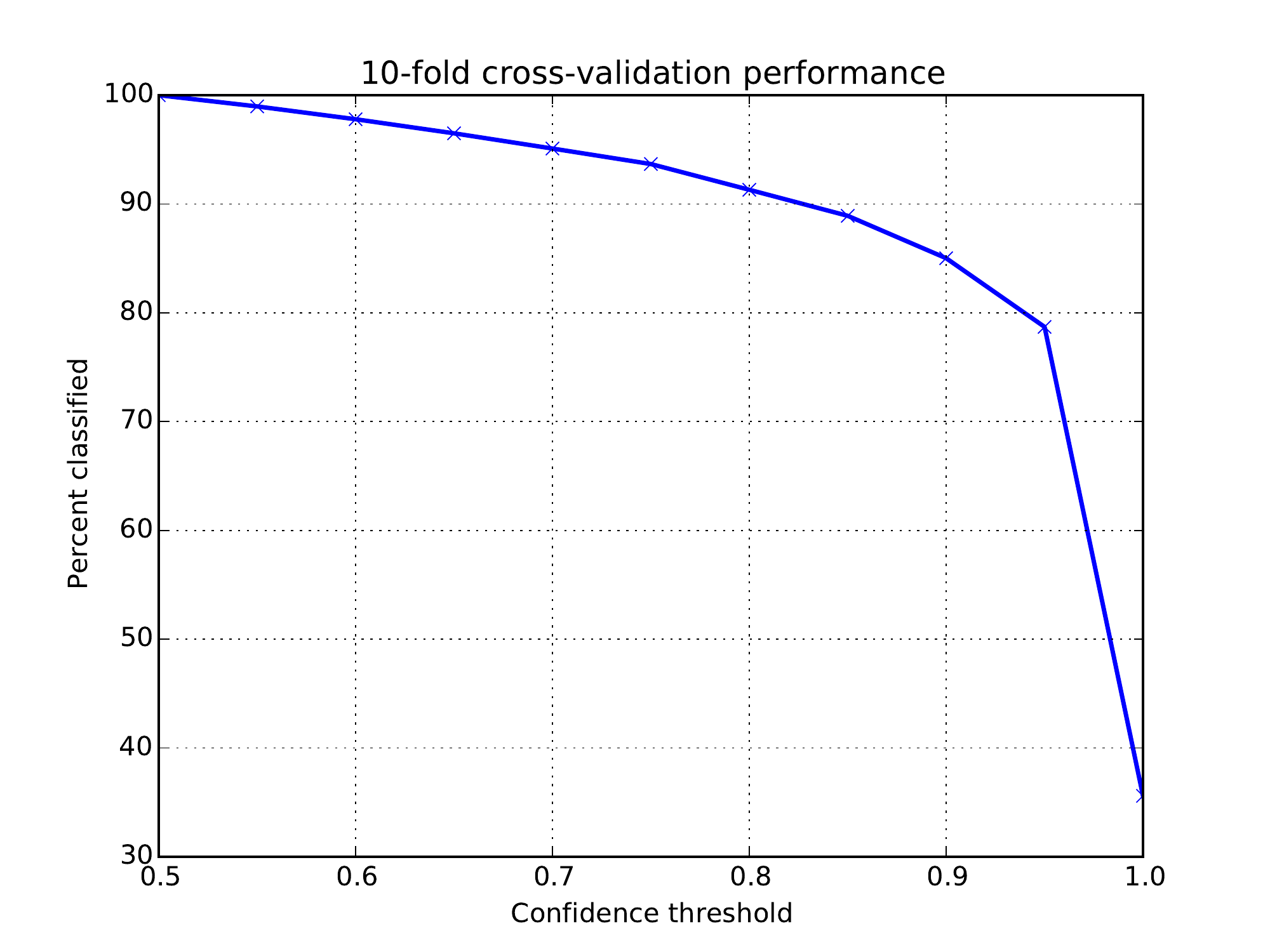}
}
\caption{V-FASTR candidate classifier performance as a function of
  confidence threshold $\tau$, using 10-fold cross-validation.}
\label{fig:perf}
\end{figure}

Our goal was a classifier with no more than a 5\% (i.e., 0.05) false
positive rate.  The classifier exceeded this requirement.  However, we
wanted to improve (reduce) the false negative (miss) rate as well.
If we impose a confidence threshold $\tau$, the classifier only
generates predictions if they have a posterior confidence $\geq
\tau$.  By default, $\tau=0.5$ for this two-class classifier, which
means that all predictions are used, since at least one class has a 
posterior probability $>= 0.5$.  Employing a higher value for $\tau$ 
increases accuracy (see Figure~\ref{fig:perf}(a)) but reduces
the number of predictions made (see Figure~\ref{fig:perf}(b)).  For
example, the rightmost data point on both plots shows that the
classifier achieved 100\% accuracy on its top 36\% most confident
predictions.  For a confidence threshold of 0.9, the classifier
achieved 98.6\% accuracy while classifying 79\% of the candidates.
This is 3.0\% higher than the original performance.  

The new confusion matrix is shown in Table~\ref{tab:cm2}.  In
comparison to the original performance obtained when making
predictions for all candidates (Table~\ref{tab:cm}), the false
positive rate decreased greatly to 0.004, and the false negative rate
halved to 0.03.  We therefore decided to
proceed with a confidence threshold of 0.9 in the operational system.
 
\begin{table}[t]
\caption{Confusion matrix given a confidence threshold of 0.9.
  Overall accuracy was 98.6\%.  Compare to Table~\ref{tab:cm}.}
\label{tab:cm2}
\begin{center}
\begin{tabular}{|l|r|r|r|} \hline
         & \multicolumn{2}{|c|}{True class} & \\ 
         & Artifact & Pulsar & Total \\ \hline
Artifact & 2436     &    73  & 2509  \\
Pulsar   &   16     &  4007  & 4023  \\ \hline
Total    & 2452     &  4080  & 6532  \\ \hline
\end{tabular}
\end{center}
\end{table}


\section{Candidate Review Pipeline and Collaborative Web Portal}
\label{sect:portal}

In this section, we describe the candidate review pipeline and recent
improvements to the collaborative web-based review system that was
originally described by \cite{Hart2014}.
%
This system consists of two major components: (1) a metadata pipeline
responsible for the capture, transfer, and storage of candidate
metadata annotations; and (2) a collaborative web portal which
provides analysts with a convenient, context-rich environment for
efficiently classifying candidate events.

\subsection{Candidate Metadata Pipeline}

The candidate review system implementation heavily leverages open
source software such as Apache
OODT\footnote{\url{https://oodt.apache.org/}} for managing the
metadata and the Apache
Solr\footnote{\url{http://lucene.apache.org/solr/}} fast-response
search server.  Apache OODT is an information management and
processing framework from the Apache Software Foundation recognized
for data archiving, metadata extraction, cataloging, querying, and
product retrieval.  Apache Solr is a widely used open source search
platform that provides rapid querying and facet-based search.

We employ the Catalog and Archive Service (CAS) Crawling
Framework~\citep{mattmann:cas13} to run at predefined intervals and 
automatically detect new candidate products, which triggers the
extraction and storage of the metadata in Solr using the OODT File
Manager~\citep{mattmann:cas13}.
%
This metadata includes information about the associated VLBA job's
observing parameters (which frequencies and antennas were used and
where the array was pointed), the date and time at which the candidate
began, the duration of the candidate, the estimated dispersion
measure, and more.

\subsection{Reviewer Web Portal Interface and Interaction}

\begin{figure}
\centerline{\includegraphics[width=5in]{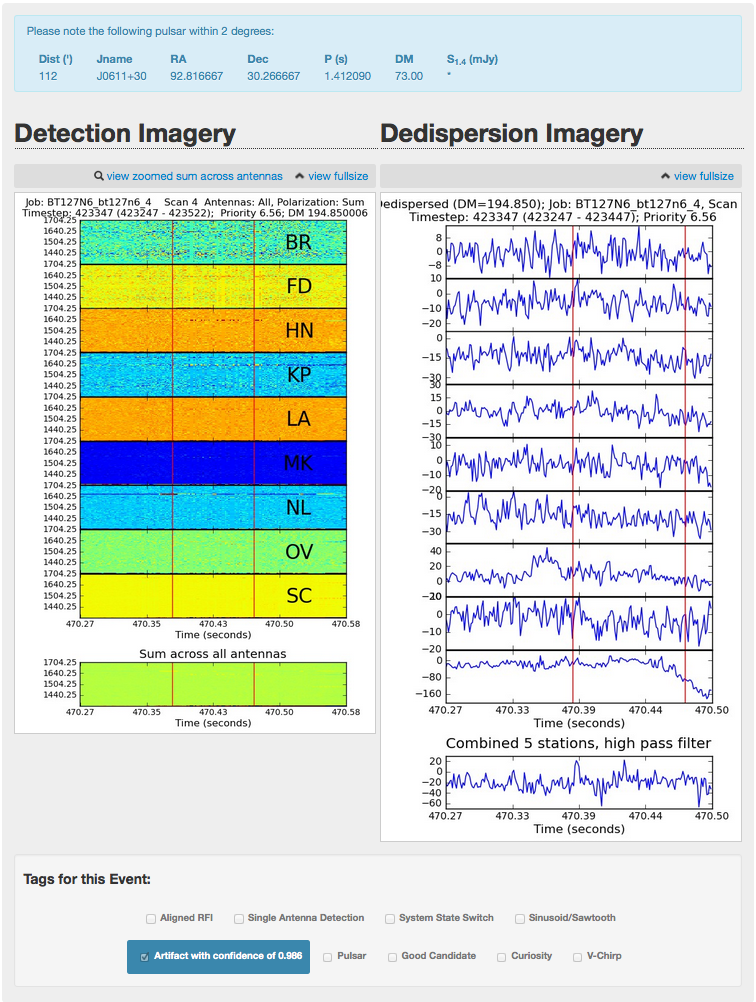}}
\caption{Web portal display for a V-FASTR candidate, including
  waterfall plots (left), de-dispersed time series (right), and nearby 
  pulsars (top).   A list of nearby pulsars appears at the top of the
  page.  Reviewers can tag the candidate using the checkboxes at the 
  bottom.  The V-FASTR classifier's prediction is shown in a blue
  highlighted box.}
\label{fig:ex2}
\end{figure}

The V-FASTR real-time candidate detection system generates key
products needed for the review and manual classification of new
candidates.  These include (1) a ``waterfall'' plot that shows signal
strength as a function of observing frequency and time, at each active
antenna, and (2) the de-dispersed time series for each active antenna,
using the DM value estimated by the candidate detection system.

The collaborative candidate review web portal provides the
geographically distributed V-FASTR science team with the ability to
quickly examine candidates as they are detected.  
Figure~\ref{fig:ex2} shows the reviewer interface to the web portal
display for an example candidate event. 
The webpage shows the waterfall plots (left) and de-dispersed time
series (right).  Vertical red lines indicate the start and end of the
event.  The portal also reports nearby pulsars (top) as an aid to
reviewers, who then have the option of tagging the event using the
checkboxes at the bottom.  

Reviewers can select from a predefined list of ``tags'' to specify the
candidate's classification (see bottom of Figure~\ref{fig:ex2}).  The
V-FASTR classifier's prediction is shown in a blue highlighted box,
along with its posterior confidence.  If the confidence is less than
0.9, the classifier's prediction is not stored, used, or displayed.
Reviewers can agree with the classifier, select another tag to correct
the classifier, or do nothing.  Any new tags created by reviewers are
stored in the Solr database.
When all of the candidates for a given job have been reviewed,
the job is archived.  The associated baseband and filterbank
data are deleted, except for the filterbank data associated with detected
candidates and the baseband data for any candidates marked as ``save''
by the reviewer (not shown in Figure~\ref{fig:ex2}).

Once a day, the V-FASTR classifier checks for new reviewer tags.  If
any are present, the classifier re-trains on all labeled data.  The
updated classifier then generates new predictions for all candidates
in the database and saves all predictions with confidence greater than
0.9.  In this way, the classifier quickly adapts to any corrections or
new information provided by the reviewers.  The web portal always
displays the latest classifier predictions.  This process is
illustrated on the right side of Figure~\ref{fig:pipeline}.




\subsection{New Review Portal Features}

\begin{figure}[t]
\centerline{\includegraphics[width=5in]{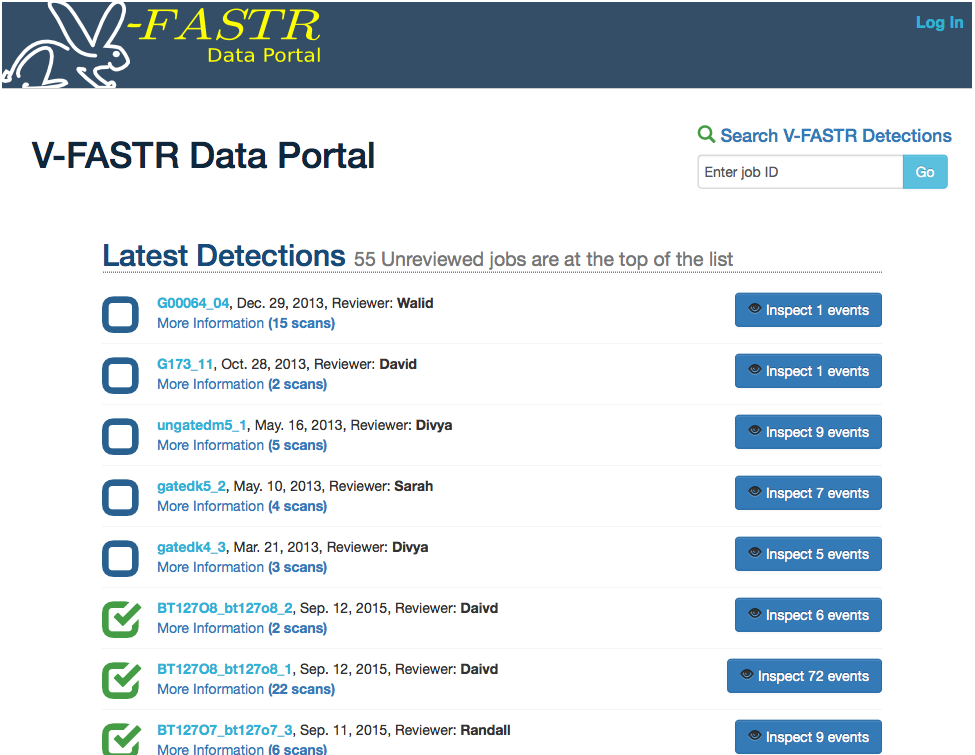}}
\caption{The V-FASTR web portal interface showing all unreviewed jobs,
  followed by reviewed jobs, in reverse chronological order.}
\label{fig:portal}
\end{figure}

We have added several new features and capabilities to the V-FASTR web
portal.  First, we greatly improved its speed and responsiveness by
upgrading the system architecture to Solr version 4.5.1,
which allows the portal to read and write directly to 
the Solr database.  This improved the speed with which new meta-data,
such as new tags created by reviewers, is stored.

Second, we redesigned the portal entry page with a more functional and
user-friendly interface (see Figure~\ref{fig:portal}).  This page
displays a list of the latest jobs that clearly distinguishes those
that have been reviewed (green checked box) from those that need to be
reviewed (blue unchecked box).  Unreviewed jobs appear first on the
list.  Previously, jobs were sorted only by date, and reviewers
sometimes had to click through several pages to find the first
unreviewed job.  Jobs are paginated with 25 jobs per page;
Figure~\ref{fig:portal} shows page 6 of the results to focus on the
boundary between the 55 unreviewed jobs and the beginning of the
reviewed jobs.

Each job has an assigned reviewer, but any reviewer can review any
job, so that the team can accommodate periods when a reviewer is sick,
on travel, or otherwise unable to review his or her assigned jobs.
Although any reviewer can review any job, it is important that there
is a nominal assignment of responsibility to ensure that no 
candidates fall through the cracks. 
Each job also has an associated button that indicates the number of
candidate events that were found within the job, as a preview of the
number of candidates that need to be reviewed.

\begin{figure}[t]
\centerline{\includegraphics[width=5in]{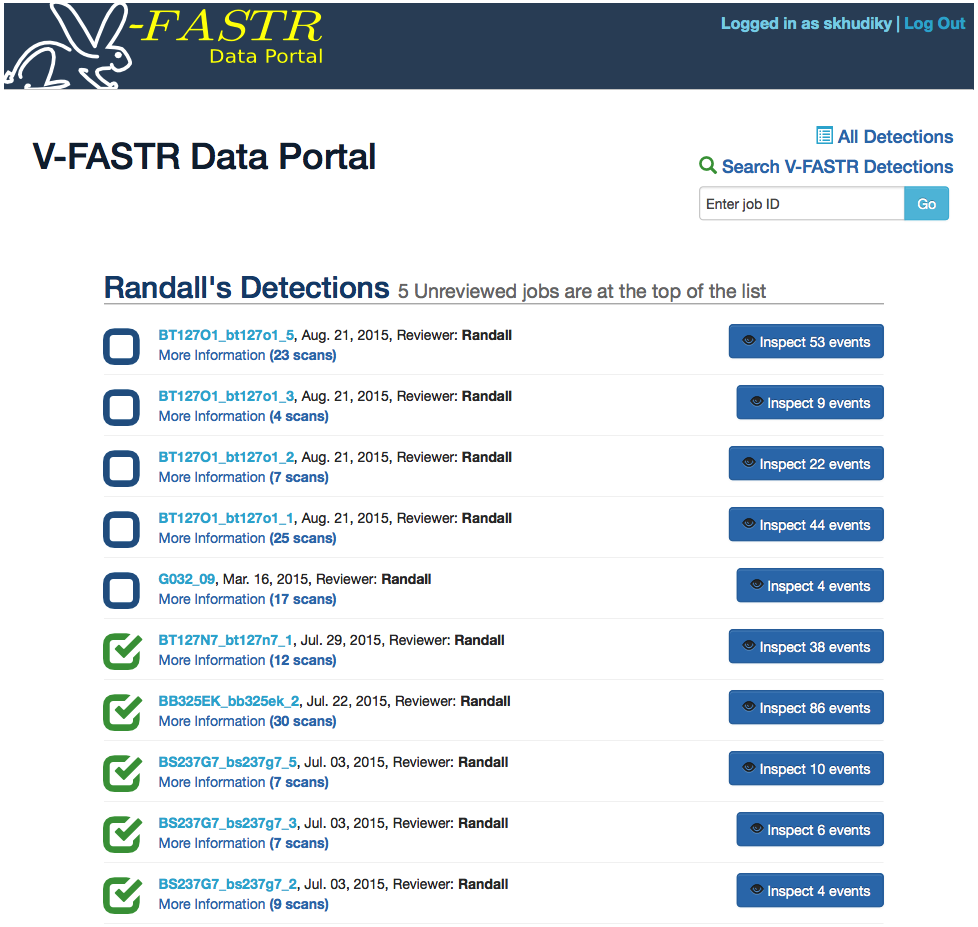}}
\caption{The V-FASTR review portal for a particular reviewer, after
  logging in.  Each reviewer is shown only those jobs assigned to to
  him or her (compare to Figure~\ref{fig:portal}).  Access to the
  full list of all jobs is available from the ``All Detections'' link
  so that reviewers can assist other reviewers with their assigned
  jobs.}
\label{fig:user-portal}
\end{figure}

Third, we added an authentication component so that reviewers can
access a customized list of jobs that contains only those jobs for
which they are the assigned reviewer.  For example,
Figure~\ref{fig:user-portal} shows the custom review list that
reviewer Randall would see.  Only jobs assigned to Randall are shown,
allowing him to quickly process his jobs.  If time permits, he can
then click ``All Detections'' to switch to the full list of all jobs
and then browse, or review, any jobs that are available.
%
User authentication now allows us to record the author of each tag. 
%
This meta-data allows us to generate statistics about per-reviewer
activity.  

The addition of authentication allowed us to open up the V-FASTR web
portal to the public.  Only logged-in users are allowed to tag
candidates or review jobs, but anyone can browse the V-FASTR jobs and
candidates in a read-only mode.  This allows other VLBA science PIs
and the interested public to see what V-FASTR has detected. 
Each candidate has its own URL, so reviewers and other users can
easily share candidates of interest and discuss them collaboratively.
This feature has been utilized by other (non-V-FASTR) VLBA users 
who are interested in time domain data to inspect potential fast
transient sources associated with their observations (e.g.,
K. Bannister, VLBA project code BB325).
  

Fourth, we modified how the candidates are shown within each job's
review page.  Previously, candidates were sorted by the time at which
they occurred.  We have now integrated the classifier's predicted tags
into the portal display so that candidates are shown in the following
order: ``good candidate'', unclassifiable, and ``pulsar.''  Candidates
classified as ``artifact'' are not shown at all, although they can be
accessed through the data search page (i.e.,~they are hidden but not
deleted).  This display significantly reduces the number of candidates
to review and prioritizes those most likely to be of interest at the
top of the page.  We employed a two-week testing and evaluation period
to assess whether this major change to candidate display was desirable
and useful.  The review team confirmed its utility, and this mode is
now the default sort ordering.

We also determined that there were other desired sorting options.  We
added an option by which users can switch between sorting by
classifier prediction (as above), SNR (strongest signals first), or by
dispersion measure (highest DM first).  All three modes have value.



\section{Results}
\label{sect:results}

We have tracked and evaluated the impact of these improvements to the
V-FASTR system in terms of reviewer time saved and a reviewer opinion
survey.  

\subsection{Reviewer Time Saved}

\begin{figure}[t]
\centerline{\includegraphics[width=5in]{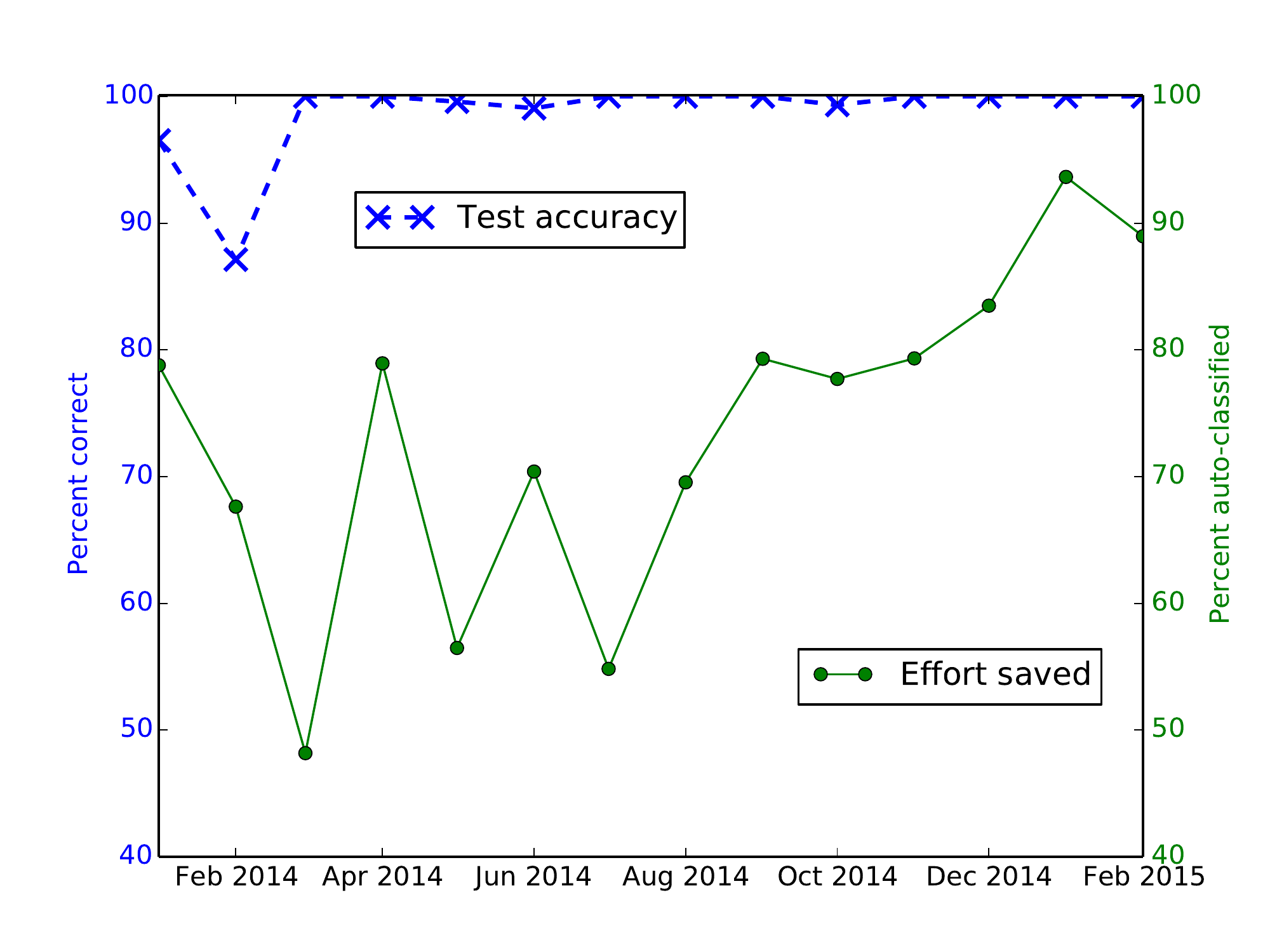}}
\caption{Retrospective study showing the candidate classifier
  performance as a function of time.  For each month, we trained a
  classifier on all preceding labeled data.  The accuracy of that
  classifier's blind predictions on new candidates observed that month
  \edit{was 99-100\% (}{is shown by the} blue dashed line. The
  reviewer effort that would be 
  saved by using the classifier's predictions to filter the candidates
  (auto-classify all ``artifact'' and ``known pulsar'' candidates) is
  shown in green.  }
\label{fig:retro}
\end{figure}

First, we conducted a retrospective evaluation of the classifier's
performance and potential benefit using the data and candidates that
were collected from January 2014 to February 2015.  The classifier was
not actually in use for filtering candidates during this period, but
the data collected during that period allowed us to determine how it
would have performed.  Starting with January 2014, at the start of
each month, we trained a classifier on all labeled candidates obtained
prior to that month.  We then used that classifier to generate blind
predictions for new candidates observed during that month.  Predictive
accuracy started at 97\% and quickly climbed to 99-100\% (see
Figure~\ref{fig:retro}, blue dashed line).

We also measured the amount of reviewer effort that would be saved by
using the classifier's predictions to filter out ``artifact'' and
``known pulsar'' candidates, so that the reviewers only needed to
examine ``good candidates'' and unclassifiable candidates.  The
fraction of candidates confidently filtered by the classifier varied
between 48-79\% from January to September of 2014 and then began to
rise, reaching 89\% by February 2015.  This indicates that the
posterior confidence estimates of the classifier improved as more
labeled candidates were available for training, enabling the confident
classification of more candidates.  Ultimately, human reviewers needed
only to examine $\sim$10\% of the incoming candidates, providing a major
time savings.

\begin{figure}[t]
\centering
\subfigure[Pulsar]{\includegraphics[width=3.2in]{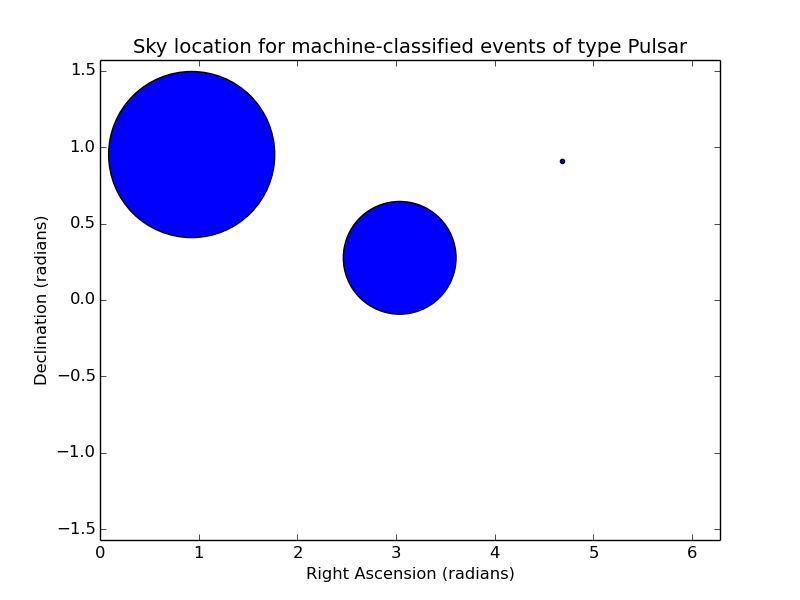}}
\subfigure[Artifact]{\includegraphics[width=3.2in]{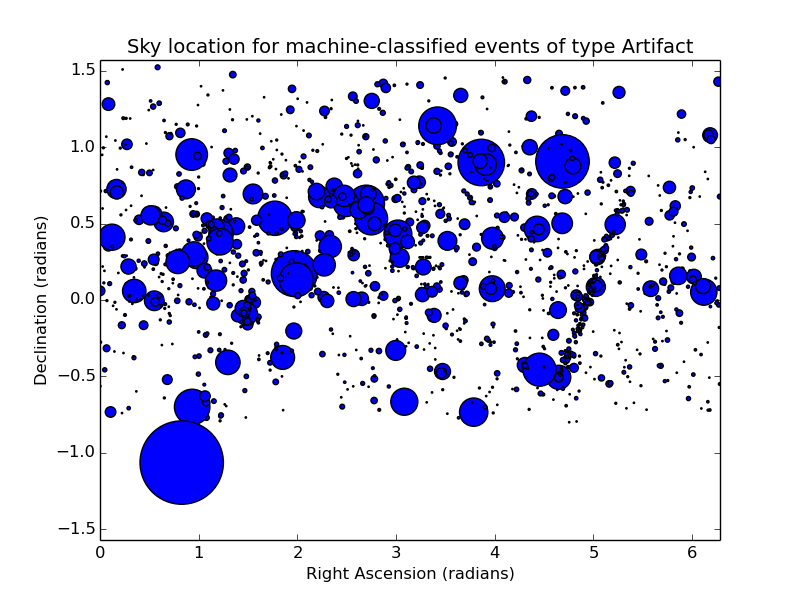}}
\caption{Sky location of candidates automatically classified as pulsar
  pulses or artifacts. The size of each circle is proportional to the
  number of automatically classified candidates found when the array
  was pointed at that sky location.}
\label{fig:det}
\end{figure}

At the time of this writing, V-FASTR has collected 197,934 candidates.
Of these, humans have labeled 4,632 pulsar pulses and 3,108 artifacts.
We applied the classifier to the 102,147 candidates detected since
January 1, 2014, from which it confidently labeled 20,096 pulsar
pulses and 45,066 artifacts.
The sky distribution of these candidates is shown in
Figure~\ref{fig:det}.  Note that these are the locations of the array
pointing center, not the candidates themselves.  The pulsar pulses
originate from three primary locations.  VLBA PIs frequently use
pulsars as calibration sources or as objects of study, so it is not
surprising that the same pulsars would be re-visited by the array.  In
contrast, artifacts appear at diverse pointings all over the sky.
There is a concentration of artifacts located along the galactic plane
(diagonal line near RA 4.5-5.5, DEC -0.5-0.6) which is likely due to
the popularity of this area as an observing location rather than any
inherent increase in RFI at that pointing.

The V-FASTR classifier has also identified 28 good candidates, which
have received special scrutiny.  In each case so far, these candidates
were determined to originate from a known pulsar location, but with an
estimated DM that was quite different from the known value of the
source pulsar.  It is possible that these are FRBs from a source that
is very near a known pulsar, but more likely that these particular
pulses exhibit some variation or RFI corruption that causes the system
to mis-estimate the DM value.  However, these unusual events are
precisely the ones that merit careful human attention, and the time
spent evaluating them is therefore well spent.


\subsection{Reviewer Survey Results}

The preceding results show that the V-FASTR classifier can greatly
reduce reviewer effort while maintaining high reliability.  We also
conducted surveyed the reviewers to assess the utility of the V-FASTR
classifier and web portal improvements from their perspective.

We conducted two surveys, one in September 2013 and the second in
September 2015.  The first survey established a baseline, using the
original web portal, against which we could compare the later
experiences after the improvements described in this paper were
implemented.  There were a total of six reviewers that participated in
the surveys.

Our primary finding is that reviewers reported spending less time
reviewing.  The average self-reported time spent reviewing decreased
from 32 to 16 minutes per week.  As described above, the number of
candidates presented to the reviewers has decreased greatly (to
10-20\% of the total number).  However, this is somewhat conflated
with a decrease in the total number of jobs going through the system
since March 2015 due to a policy change.

We also found that the self-reported time spent per candidate {\em
  increased}, from an average of 5.4 to 10.2 seconds.  This suggests
that the candidates being reviewed require additional thought and are
potentially more ambiguous or more interesting to the reviewer.  When
asked to report the percent of candidates that the reviewers
considered (subjectively) interesting, the average increased slightly
from 0.9\% to 1.2\%.  Because interesting candidates are rare, it is
not surprising that this percentage is low.  However, we aim to
continue increasing the fraction.


Our most recent survey culminated in a question aimed specifically at
assessing the impact of the incorporation of the classifier into the
system.  We asked: ``Has the incorporation of the classifier, which
now hides candidates tagged as `pulsar' or `artifact', reduced the
time/effort you invest in reviewing?''  As shown in the following
table, we found that 4 of 6 reviewers reported benefiting from the
classifier's decisions.

\begin{center}
\begin{tabular}{|l|p{5in}|} \hline
Votes & Response \\ \hline
2 & Yes. I spend much less time (or effort) in reviewing each job. \\
2 & Yes. I spend a little less time (or effort) in reviewing each
job. \\
0 & No. I spend about the same amount of time (or effort) in reviewing
each job, compared to the previous system with no classifier-based
filtering. \\
2 & I'm not sure; I haven't had enough recent jobs to review to see a
difference. \\
\hline
\end{tabular}
\end{center}

\section{Conclusions and Future Work}
\label{sect:conc}

In this paper, we have presented two major improvements to the V-FASTR
fast transient detection and review system.  First, we incorporated a
machine learning classifier to automatically filter out known types of
events (pulsar pulses and artifacts) and enable human reviewers to
devote their time to the most promising candidates.  This classifier
has a 98.6\% accuracy on historical data and a 99-100\% accuracy on
(labeled) newly collected data.  It retains only predictions of at
least 0.9 confidence. 
This reduces the reviewer workload by 80-90\%; only $\sim$10\% of the
candidates need human eyes.  The remaining candidates can still be
accessed through the V-FASTR archive, enabling the compilation of
statistics in terms of candidate sky location, DM, SNR, etc.

Second, we implemented several efficiency and usability improvements
to the V-FASTR web review portal that has streamlined the review
process and made it possible for us to open the V-FASTR archive to the
public. 

These improvements have enhanced the ability of the V-FASTR science
team to quickly review and tag candidate transient events detected
during commensal processing of VLBA observations.
A similar approach could be used in the future to manage the large
volume of interesting candidates that are anticipated to be collected
by the Square Kilometre Array and other future instruments.

\acknowledgments

{\it Facilities:} \facility{VLBA}, \facility{NRAO}.  

We would like to thank Benyang Tang for his original version of the
V-FASTR candidate classifier, Luca Cinquini and Andrew Hart for the
initial web portal and data crawler architecture design and
implementation, Tyler Palsulich for integrating the pulsar database
into web portal, and Sarah Burke-Spolaor, Walter Brisken, and Walid
Majid for contributing labeled data and feedback on the classifier and
portal system.

The International Center for Radio Astronomy Research is a joint
venture between Curtin University and The University of Western
Australia, funded by the State Government of Western Australia and the
joint venture partners. Steven J.~Tingay is a Western Australian
Premiers Research Fellow.
This work was done in part at the Jet Propulsion Laboratory under a
Research and Technology Development Grant, under contract with the
National Aeronautics and Space Administration.  Copyright 2015.
All Rights Reserved. US Government Support Acknowledged.




\clearpage

\bibliographystyle{apalike}
\bibliography{references}

\clearpage

\end{document}